\documentclass[11pt,letterpaper]{article}
\usepackage{amsmath,amssymb,amsfonts}
\usepackage{graphicx,wrapfig}
\usepackage[usenames,dvipsnames]{xcolor}
\usepackage{skak}
\usepackage{caption}
\usepackage[affil-it]{authblk}

\graphicspath{{./figures/}}
\numberwithin{equation}{section}
\allowdisplaybreaks

\newcommand{\be}{\begin{equation}}
\newcommand{\ee}{\end{equation}}
\newcommand{\ba}{\begin{aligned}}
\newcommand{\ea}{\end{aligned}}
\newcommand{\ben}{\begin{eqnarray}\displaystyle}
\newcommand{\een}{\end{eqnarray}}

\newcommand{\cE}{\mathcal{E}}
\newcommand{\cF}{\mathcal{F}}

\newcommand{\cI}{\mathcal{I}}

\newcommand{\cN}{\mathcal{N}}
\newcommand{\cO}{\mathcal{O}}

\newcommand{\cS}{\mathcal{S}}
\newcommand{\cT}{\mathcal{T}}

\setboardfontsize{8}

\begin{document}

\begin{titlepage}

\title{Relations between supersymmetric structures in UV and IR for $\cN=4$ bad theories}

\author{Denis Bashkirov
  \thanks{Electronic address: \texttt{dbashkirov@perimeterinstitute.ca}}}
\affil{Perimeter Institute for Theoretical Physics\\
Waterloo, Ontario, ON N2L 2Y5, Canada}

\maketitle

\abstract{We investigate options for the structure of the infrared fixed points of $\cN=4$ bad theories in three dimensions. Unitarity constraints allow a number of possibilities, not necessarily a product of an interacting $\cN=4$ SCFT and free theories. For each option we provide relations between the UV and IR $R-$symmetry groups. For some of them we give examples. In particular, the $\cN=4$ $SU(2)$ SYM with two fundamental hypermultiplets is an example of a bad theory which flows to an interacting irreducible SCFT in the IR. The question of whether all the options are realized remains open.}

\end{titlepage}

\tableofcontents

\section{Introduction}

One of the most interesting problems in QFT is understanding the IR behavior of asymptotically free theories. In three space-time dimensions all gauge theories are asymptotically free and, hence, strongly coupled in the infrared. Unfortunately, nonsupersymmetric theories are essentially intractable as long as their IR behavior is concerned. However, there has been a significant progress in recovering certain information about IR limits of supersymmetric $\cN=2$ gauge theories in three dimensions. It includes exact computations of partition functions on certain space-time backgrounds \cite{KWY,Kim,Jafferis,IY} and conformal dimensions of BPS operators \cite{Jafferis} for cases when accidental IR symmetries do not get in the way.

For ``bad" theories IR accidental symmetries do get in the way, and it is not known how to compute these quantities, and there is very little information about the IR fixed point.

The term ``bad theory" for $\cN=4$ supersymmetric three dimensional theories was introduced in \cite{GW} for theories whose UV $R-$symmetry group $SO(4)_r\equiv SU(2)_X\times SU(2)_Y$ is not  the $R-$ subgroup $SO(4)_R$ of the superconformal group of the IR fixed point. As in four dimensional QFT's \cite{Seiberg}, an indicator of ``badness" is presence of local operators in UV theory whose conformal dimensions at the IR fixed point violate unitarity. For BPS operators in a SCFT, the conformal dimensions are equal to their $R-$charges. Unless a UV BPS operator becomes $Q-$exact in the IR, this allows to identify at least some of bad theories.

The analysis of $R-$charges of monopole operators \cite{BKW, BKW2} done in \cite{GW} identified some bad theories. For example, it was found that an $\cN=4$ super Yang-Mills with gauge group $U(N_c)$ and $N_f$ hypermultiplets in fundamental representation is bad when $N_f\le 2N_c-2$.

With the development of the localization technique allowing to compute partition functions of supersymmetric theories on a class of three-manifolds \cite{KWY,Kim, IY} it becomes possible to introduce another indicator of a bad theory. Namely, the path integral of a background gauged $R-$supermultiplet on ${\mathbb S}^2\times {\mathbb S}^1$ computes a partition function of the UV theory which becomes the superconformal index \cite{BBMR} of the IR theory when the $R-$symmetry is the superconformal $R-$symmetry. The superconformal index of a SCFT should be a finite function of fugacities. Thus, a divergent result indicates that the UV theory is bad. The correlation between divergence of the partition function of the theory on ${\mathbb S}^3$ and ``badness" of the theory was noted in \cite{WY}.

We only consider $\cN=4$ and $\cN=8$ supersymmetric UV theories as there are no superconformal $\cN=3,5,6,7$ theories with relevant perturbations preserving the amount of supersymmetry. Hence there are no bad $\cN=3,5,6,7$ UV theories. Theories with a less amount of supersymmetry ($\cN=2,1,0$) are not manageable at present.

As far as we know, it is generally assumed that a bad UV theory becomes in IR a product of an IR limit of a good theory and a free theory. This assumption stems from the expectation that for bad theories there is no point on the moduli space which preserves all the UV symmetries, so that the scaling limit produces In\"on\"u-Wigner contraction of some of the UV symmetries which become translations along some directions in the IR leading to free fields \cite{GMT}. Although this is indeed what happens in the examples of bad theories \cite{Itamar,GK}, there is no theorem stating that this must be the case\footnote{I thank Davide Gaiotto for clarification of this point.}. In section 2 we start with a discussion of a motivating counterexample where the IR limit of a bad $\cN=4$ theory is an irreducible SCFT without any free fields. Although this theory is a very special type of $\cN=4$ theory being actually an $\cN=8$ supersymmetric, this example raises the question of whether the nature of IR limits of bad theories can be more diverse than previously assumed.

In this paper we take the first step towards the answer to this question which is the analysis of possible relations between UV and IR $R-$symmetry groups and their correlation with ``compositeness" and existence of free sectors in the IR. It turns out that some variants are not allowed by unitarity, yet, many others are allowed.

This analysis can also have a more practical application for weeding out candidates for the IR limits of a bad UV theory. Indeed, given a UV bad theory and a number of candidates for its IR fixed point, one way to pick up the correct one, or at least to reduce the number of candidates is to compute partition functions using the UV and IR Lagrangians gauging UV $R-$symmetries and to compare them as was recently done for partition function for some bad theories on ${\mathbb S}^3$ in \cite{Itamar}. But to do this one needs to know how the $R-$symmetries are related. That is, what symmetries of the IR theory correspond to the $SO(4)_r$, for example. We provide a classification of all these relations as functions of the types of IR fixed points.

In section 3.1 we give some examples of relations which are ruled out by unitarity. In section 3.2 we list all superconformal multiplets which contain primary conserved currents. In section 3.3 we consider the simplest bad $\cN=4$ theory -- the $U(1)$ gauge theory to pave the way for discussions in subsequent section. In sections 3.4 and 3.5, using results from section 3.2, we discuss all options for the structure of the IR fixed points of $\cN=4$ bad theories compatible with unitarity. These are generated by

\begin{itemize}
\item The IR fixed point is an $\cN=4$ irreducible interacting SCFT with flavor symmetry group $G$ including at least one factor of $SU(2)$.
\item The IR fixed point is the free $\cN=4$ SCFT.
\item The IR fixed point is a product of at least two interacting $\cN=4$ SCFT's without any flavor symmetries.
\item The IR fixed point is the free $\cN=8$ SCFT.
\end{itemize}
The IR fixed point of an arbitrary bad $\cN=4$ theory is some product of these sectors.

Finally, in section 4 we consider an interacting $\cN=4$ SCFT whose UV description is bad in some detail.

\section{$\cN=8$ super YM with gauge groups $SU(N)$}

Consider maximally supersymmetric super Yang-Mills theories in three dimensions with gauge groups $SU(N)$. They are just dimensional reductions of $\cN=4$ SYM theories in four dimensions with the same gauge groups. The matter content is the following. There are seven real scalar fields in the adjoint representation of the gauge group, six of which come from the six scalars in four dimensions, and the seventh comes from the component of the gauge field along the reduced direction. As in four dimensions, there are four ``Weyl" fermions which become eight Majorana spinors in three dimensions. The $R-$ symmetry group is $SO(7)_R$ with scalars in the vector representation ${\bf 7}$, and fermions in the real spinor representation ${\bf 8}$. The supercharges are in the representation $({\bf 8, 2})$ of $Spin(7)\times SO(3)$. The  $SO(3)$ factor is the Lorentz group.

These theories are particular examples of $\cN=4$ supersymmetric theories. The $R-$group $SO(4)_R$ corresponding to $\cN=4$ supersymmetric structure is embedded in $SO(7)_R$ in the following way. $Spin(7)_R$ is naturally decomposed into $Spin(7)_R\supset Spin(4)\times SU(2)=SU(2)_1\times SU(2)_2\times SU(2)$. The $Spin(4)_R$ of the $\cN=4$ structure is then $Spin(4)_R=SU(2)_2\times SU(2)$. It is easy to find the irreducible representations of $Spin(4)_R$ into which the matter fields fall.

\begin{align}
{\bf 7}={\bf 4\times1}+{\bf 1\times 3}=({\bf 2,1})+(\bf{1,3}),\\
{\bf 8}={\bf 2\times 2}+{\bf 2'\times 2}=({\bf 1,2})+({\bf 2,2})
\end{align}

The scalars $(\bf{1,3})$ together with fermions $({\bf 2,2})$ and gauge fields form the $\cN=4$ vector multiplet, and scalars $({\bf 2,1})$ with fermions $({\bf 1,2})$ form a single $\cN=4$ hypermultiplets.

In the $\cN=2$ language the supermultiplets are: one vector multiplet, one chiral multiplet with $R-$charge $R=1$ and two chiral multiplets with $R-$charges $R=1/2$.

Correspondingly, $R$-charges of BPS monopole operators with GNO charges $H=(n_1,...,n_N)$ subject to the condition $\sum_{i=1}^Nn_i=0$ are
\begin{align}
R=-\sum_{\alpha\in\Delta_+}|\alpha(H)|+\frac{1}{2}\sum_{\rho}|\rho(H)|=-\sum_{\alpha\in\Delta_+}|\alpha(H)|+\frac{1}{2}\sum_{\alpha\in\Delta}|\alpha(H)|=0
\end{align}

The first sum runs over all positive roots, the second one runs over all roots (which for the adjoint representation coincide with the set of weights) which is twice the sum over positive roots. The $R=1$ chiral multiplets do not contribute to the $R-$charge of monopole operators \cite{BK}.

By the argument of \cite{GW} BPS operators having zero $R-$charge imply that the $R-$symmetry is not the one which enters the superconformal algebra for the IR fixed point, because the conformal dimension of a BPS operator equals its $R-$ charge (minus $R-$ charge for an anti-BPS operator), and in a unitary conformal theory the lowest allowed conformal dimension of a local operator is $\Delta=1/2$ in three dimension. The only exception is the unit operator $\widehat 1$ whose conformal dimension is zero.

The conclusion that the UV $R-$ charge is not one appearing in the $\cN=2$ superconformal structure of the IR limit of the $\cN=8$ SYM can alternatively be reached by considering the 'superconformal' index which is computed by the path integral with the UV theory put on ${\mathbb S}^2\times{\mathbb S^1}$ appropriately twisted along the circle \cite{Minwalla,Kim,IY}.
There are different ways to put a supersymmetric theory on a curved space-time background preserving some supersymmetry. They are parameterized by the choice of the supermultiplet containing the stress tensor \cite{CDFK}. This supermultiplet is weakly gauged, and some of the background gauge fields (including the Levi-Civita connection, equivalently, the metric) are fixed to nonzero values.

To compute the superconformal index one uses the $R-$multiplet containing the superconformal $R-$ current, supercurrents and the stress tensor, among other fields. If the superconformal $R-$current is not correctly identified in the UV, the path integral computes a certain partition function but not the superconformal index.

This is what happens in the case in hand. The partition function is not the superconformal index because it is infinite. The cause of divergence is the fact that the contribution of every summand parameterized by the GNO charge contains a piece which is the same for all GNO charges. This happens because every bare monopole operator/state with a nonzero GNO charge does not contribute to the $R-$charge and energy (in radial quantization).

In all examples of bad theories where the IR limit was determined, it was found that the IR theory was a product of a good theory and a collection of free theories. In the present case, which is a particular kind of $\cN=4$ theory we show that there is a single interacting irreducible theory.

First of all, let us recall what is known about the IR limit of $\cN=8$ SYM. It is an $\cN=8$ superconformal theory with $R-$ symmetry group $Spin(8)_R$.

To show that for the gauge group $SU(N)$ there is a single sector in the IR we use the superconformal index. Although, as we argued above, it cannot be computed from the UV theory, we can use a trick. The $U(N)$ maximally supersymmetric Yang-Mills theory is known to be equivalent to the ABJM theory \cite{ABJM} with gauge group $U(N)\times U(N)$ and Chern-Simons levels $(1,-1)$ whose superconformal index can be computed to a finite degree in fugacity $x$ without a problem. Furthermore, the $U(1)$ $\cN=8$ SYM is dual in the IR to the free $\cN=8$ superconformal theory of four chiral multiplets whose index can be computed as well. Then the superconformal index of the $SU(N)$ gauge theory is
\begin{align}
{\cI}_{SU(N)}(x)=\frac{\cI_{ABJM}(x)}{\cI_{free}(x)}.
\end{align}

Such a computation was performed in \cite{B} for the IR limit of $SU(2)$ theory and the result was found to be
\begin{align}
{\cI}_{SU(2)}(x)=1+10x+\cO(x^2)
\end{align}

It was explained in the same paper that absence of terms $x^{1/2}$ implies that the theory is not free, and the coefficient $10$ in front of $x$ implies that it is irreducible. If, for example, the coefficient were $20$, this would mean the theory is composed of two sectors, each with $\cN=8$ superconformal symmetry.

It is very easy to compute the superconformal index for ABJM theories with arbitrary $N$ to the first order in $x$. This gives the same answer for $SU(N)$ as for the $SU(2)$ case. The conclusion is the same: the IR limit is an irreducible interacting $\cN=8$ superconformal theory. That the IR theory is irreducible is also clear from the moduli space.

It is easy to understand why no $U(1)_R$ symmetry in the IR corresponding to an $\cN=2$ superconformal subalgebra is an UV $R-$symmetry. $Spin(7)_R$ is enhanced in IR to $Spin(8)_R$ with seven accidental currents which form a fundamental representation of $Spin(7)_R$ which together with the $Spin(7)_R$ currents make up the adjoint representation of $Spin(8)_R$. In this process the eight supercharges $Q$ which were in the spinor representation of $SO(7)_R$ become a spinor representation of $SO(8)_R$. However, in the superconformal $\cN=8$ superalgebra, the eight supercharges must be in a vector representation of $SO(8)_R$. This means that to make up the superconformal structure, we must do a triality automorphism of $SO(8)_R$ to turn the spinor representation into the vector representation.

An important thing is that when we consider a $U(1)$ subgroup of $SO(8)_R$ whose commutant is $SO(6)$ after using the triality transformation, that is, when we single out an $\cN=2$ superconformal algebra of the $\cN=8$ superconformal algebra, the current corresponding to the $U(1)_R$ will necessarily contain an ``accidental" current which is conserved only at the IR fixed point, but not in the UV. This explains why no $U(1)-$ current of $Spin(7)_R$ is a superconformal $R-$current.

This process can be checked in detail for the $U(1)$ $\cN=8$ theory which is dual to the theory of four free chiral fields. For details see appendix A.

Thus, maximally supersymmetric $SU(N)$ gauge theories provide examples of bad theories whose IR limits are irreducible interacting $\cN=8$. One way to view this resolution is  as being due to the triality automorphism of $SO(8)$. Another way, when we consider the UV theory as an $\cN=4$ supersymmetric QFT, is the plethora of $SU(2)$ global symmetry groups in the IR theory. This raises the question whether there are other $\cN=4$ bad theories which do not contain free sectors in IR due, perhaps, to numerous $SU(2)$ global symmetry groups in IR. We investigate this possibility in the remaining sections.

\section{Classification of bad $\cN=4$ theories}

\subsection{Nontriviality of unitarity constraints}

The first step in classification of bad theories in terms of their IR behavior is the classification of all possible relations between the UV and IR $R-$symmetry group. As we will see, the requirement of unitarity and superconformal structure in IR impose nontrivial constraints on these relations, killing some of the possibilities, but at the same time they leave the following possibilities:

\begin{itemize}
\item The IR theory is an irreducible interacting $\cN=4$ SCFT whose complete flavor symmetry group $G$ contains at least one factor of $SU(2)$: $G=SU(2)\times H$.
\item The IR theory is irreducible with supersymmetry enhanced to $\cN=8$.
\item The theory is irreducible in the $\cN=4$ sense but is free.
\item The theory is reducible with each sector having at least $\cN=4$ supersymmetry. Furthermore, each sector can be free or interacting. In particular, it is possible that all sectors are interacting SCFT's.
\end{itemize}

Let us start with an example of an inconsistent relation between UV and IR $R-$symmetry groups.

Consider a theory whose group of global symmetries is $SU(2)_1\times SU(2)_2\times SU(2)_3$. By global symmetries we mean all symmetries except space-time (super)symmetries. In particular, these include superconformal $R-$symmetries and flavor symmetries. As we explain later, in free theories there are global symmetries that are neither superconformal $R-$symmetries nor flavor symmetries. Suppose that the superconformal $SO(4)_R=SU(2)_R\times\widetilde{SU(2)}_R$ symmetry group and the UV $R-$symmetry group $SO(4)_r=SU(2)_X\times SU(2)_Y$
\begin{eqnarray}
& SU(2)_R=SU(2)_1,\qquad\qquad SU(2)_X=SU(2)_2,\nonumber\\
& \widetilde{SU(2)}_R=SU(2)_2,\qquad\qquad SU(2)_Y=SU(2)_3.
\end{eqnarray}

This choice does produce a 'bad' relation between the UV and IR $R-$ symmetry groups. Indeed, the $R-$ charges of the IR theory $R$ and the UV theory $r$ are
\begin{eqnarray}
 R=\pm J\pm\widetilde{J}=\pm J_1\pm J_2,\qquad r=\pm J_X\pm J_Y=\pm J_2\pm J_3
\end{eqnarray}
  which are different. Here all $J$'s are the third projection of the spin for the corresponding $SU(2)$ group.

However, this relation is forbidden by unitarity and superconformal invariance. To see this, recall that the supercharges must be in the vector representation of $SO(4)$ $R-$symmetry groups. Furthermore, they must be in some representation of the subgroup $SU(2)_1\times SU(2)_2\times SU(2)_3$ of the full group of global symmetries. There are two options:
\begin{eqnarray}
(i)\quad Q=(2,2,2)+...,\qquad (ii)\quad Q=(2,2,1)+(1,2,2)+....
\end{eqnarray}

The ``$...$" stand for some other representation. The first option cannot be realized since it does not provide the four supercharges that are part of the superconformal algebra which must be $(2,2,1)$. The second option is more subtle. It requires existence of primary conserved currents with conformal dimension $\Delta=5/2$, spin $j=3/2$ and in representation $(1,2)$ of the superconformal $R-$symmetry group $SO(4)_R=SU(2)_R\times\widetilde{SU(2)}_R$ which upon integration over space give the supercharges. As we show below, unitarity forbids existence of a supermultiplet containing such operators.

\subsection{Classification of supermultiplets containing (super)currents}

This subsection lists the superconformal multiplets which contain conserved (super)currents. In a superconformal theory, a conserved supercurrent which gives a supercharge, is a primary local operator with spin $j=3/2$ and conformal dimension $\Delta=5/2$. It is also in some representation of the $R-$symmetry group, $SO(4)_R$ for $\cN=4$. Although it is a primary operator\footnote{That is, it is annihilated by the special conformal generators $K_\mu$.}, it may or may not be a superprimary operator\footnote{A superprimary operators is an operator annihilated by the superconformal charges $S_{\alpha}$.}. If it is not a superprimary operator, it is a superdescendent, that is, it is obtained by the action of supercharges on another local operator. Each local operator in a superconformal theory belongs to some superconformal multiplet. On the lowest level of a superconformal multiplet is a superprimary operator in some representation of the bosonic subgroup of the superconformal group. That is, it has a certain conformal dimension $\Delta$, a certain spin $j$, and belongs to a certain representation of the $R-$symmetry group $SO(4)_R$ with the highest weight $(h_1,h_2)$ where $h_1\ge h_2$ and $h_2$ are Cartan operators corresponding to rotations in the planes $12$ and $34$.

The requirement of unitarity imposes nontrivial constraints on the superconformal multiplets \cite{Minwalla} which are expressed as inequalities between quantum numbers of superprimary operators as follows

\begin{eqnarray}
& \Delta\ge h_1+j+1, \qquad\hbox{for}\quad j>0.
\end{eqnarray}

When the inequality is saturated, some operators in the superconformal multiplet acquire zero norm as defined by two-point correlation function, and can be consistently omitted which leads to shortening of the superconformal multiplet.

The spin-zero superprimary operator corresponds either to an isolated short multiplet with the conformal dimension of the superprimary $\Delta=h_1$ or to multiplets with
\begin{eqnarray}
& \Delta\ge h_1+1
\end{eqnarray}

with the extreme case $\Delta=h_1+1$ corresponding to another short superconformal multiplet.

These restrictions give the following superconformal multiplets which contain primary conserved (super)currents.

\begin{center}
   \begin{tabular}{| l | l || l | l | l | l |}
    \hline
     & Name & $\cT$ & $\cE$ & $\cF$ \\
    Operators & & & & \\ \hline\hline
    $j=0$ & $\Delta=1$ & $(1,1)$ & $(2,2)$ & $(3,1)$ or $(1,3)$\\
    $j=\frac12$ & $\Delta=\frac32$ & $\surd$ & $\surd$ & $\surd$\\
    $j=1$ & $\Delta=2$ & $(3,1)+(1,3)$ & $(2,2)$ & $(1,1)$\\
    $j=\frac32$ & $\Delta=\frac52$ & $(2,2)$ & $(1,1)$ & $-$\\
    $j=2$ & $\Delta=3$ & $(1,1)$ & $-$ & $-$ \\
    \hline
    \end{tabular}
     \captionof{table}{(Super)currents multiplets in $\cN=4$ SCFT. Part 1.}
\end{center}

\begin{center}
   \begin{tabular}{| l | l || l | l | l | l |}
    \hline
     & Name & $\Psi$ & $\cN$ & $\cS$ \\
    Operators & & & & \\ \hline\hline
    $j=0$ & $\Delta=1$ & $-$ & $-$ & $-$\\
    $j=\frac12$ & $\Delta=\frac32$ & $(1,1)$ & $-$ & $-$\\
    $j=1$ & $\Delta=2$ & $(2,2)$ & $(1,1)$ & $-$\\
    $j=\frac32$ & $\Delta=\frac52$ & $(3,1)+(1,3)$ & $(2,2)$ & $(1,1)$\\
    $j=2$ & $\Delta=3$ & $-$ & $(3,1)+(1,3)$ & $(2,2)$\\
    \hline
    \end{tabular}
     \captionof{table}{(Super)currents multiplets in $\cN=4$ SCFT. Part 2.}
\end{center}

In this table the pair of numbers $(m,n)$ denote the representation of the local operators under the superconformal $R-$symmetry group $SO(4)_R\equiv SU(2)_R\times\widetilde{SU(2)}_R$, the sign $``-"$ denotes absence of the corresponding operator, and $``\surd"$ denotes its presence. Because the operators with $j=1/2$, $\Delta=3/2$ are not conserved currents, they are of no interest to us, unless they are superprimary, so we did not show their $SO(4)_R$ quantum numbers.

Furthermore, for each multiplet only the most important operators are shown -- primary conserved currents and superprimary operators. For each supermultiplet there are additional operators which we do not show.

In addition, there is a supermultiplet whose superprimary is a ``stress tensor" -- an $R-$singlet conserved current with spin $j=2$ and conformal dimension $\Delta=3$. Such multiplets only exist in the free $\cN=4$ and $\cN=8$ theories or reducible theories containing the free $\cN=4$ of $\cN=8$ sectors. For example, in the free $\cN=4$ theory there are two $SO(4)_R$ singlet ``stress-tensors": $T^{(b)}=\phi^i\partial^2\phi^i$ and $T^{(f)}_{\mu\nu}=\psi^i\partial\psi^i$. Their sum $T^{(b)}+T^{(f)}$ is the stress tensor that lives in the stress-tensor multiplet $\cT$ and produces Poincare charges, while their difference $T^{(b)}-T^{(f)}$ is a superprimary.

The $\cT$ multiplet is the superconformal multiplet identical to the stress-tensor multiplet. Thus presence of two (or more) such multiplets in the theory implies ``compositeness" -- the superconformal theory is the product of two (or more) $\cN=4$ SCFT's\footnote{We do not discuss a more exotic possibility of a SCFT which is obtained tensoring two SCFT's and then gauging some symmetry.}.

The $\cE$ multiplet, or more precisely, a certain collection of such multiplets is present when there is an enhancement of supersymmetry from $\cN=4$ to $\cN=5$, $\cN=6$ or $\cN=8$ \cite{BK}.

The $\cF$ multiplet is the multiplet of flavor currents. It does not contain primary conserved supercurrents, but we put it in the table for completeness.

The $\Psi$ multiplet has a spinor with conformal dimension $\Delta=3/2$ as the superprimary operator.

For each of the multiplets, it is a nontrivial statement that they contain certain operators on higher levels, because although they may be allowed by group theory, they may have zero norm, and so must be omitted.

The $\Psi$, $\cN$ and $\cS$ multiplets are only present when the IR SCFT is composed of several free sectors and, possibly, several non-free ones. The details are provided in the next two sections.

\subsection{An example of a bad theory}

We start with a concrete example of a bad theory.

\begin{itemize}
\item $\cN=4$ $U(1)$ super Yang-Mills.
\end{itemize}

This theory has the following field content: $\cN=2$ free vector multiplet together with the free $\cN=2$ chiral multiplet. The global symmetry group is $SO(4)_r\equiv SU(2)_X\times SU(2)_Y$. Two real scalars from the chiral multiplet and one real scalar from the vector multiplet form the representation $(3,1)$, while four Majorana fermions belong to representation $(2,2)$.

The four real supercharges are in representation $(2,2)$ as well.

By the criteria explained above, this theory is bad. That is, $SO(4)_r\equiv SU(2)_X\times SU(2)_Y$ is not the IR superconformal $R-$symmetry group $SO(4)_R\equiv SU(2)_R\times\widetilde{SU(2)}_R$ \cite{GW}.

Let us find out the relation between the two $SO(4)$ groups.

It is known that the free $U(1)$ gauge field in three dimensions can be dualized to a compact scalar whose target space is a circle with radius $e^2$ where $e$ is the dimensionful gauge coupling constant of the gauge field. The IR limit of this theory corresponds to sending $e^2$ to infinity, that is, it is the decompactification limit for the scalar.

Returning to the case of the free $U(1)$ $\cN=4$ vector multiplet, this implies that its IR limit is the theory of a free hypermultiplet. In other words, it is the theory of four real free scalars and four free Majorana fermions. The full global symmetry of this theory is $SO(4)_b\times SO(4)_f\equiv SU(2)_b^{(1)}\times SU(2)_b^{(2)}\times SU(2)_f^{(1)}\times SU(2)_f^{(2)}$. Here $SO(4)_b$ rotates the four real scalars, and $SO(4)_f$ rotates the four free Majorana fermions.

That is, the fields are in the representations

\begin{align}
\hbox{scalars}=(2,2;1,1),\qquad\hbox{fermions}=(1,1;2,2).
\end{align}

Spin-$3/2$ conserved currents are obtained by taking products of fermions and space-time derivatives of scalars which gives representation $(2,2;2,2)$ of the global symmetry group.

The superconformal $R-$symmetry group is $SU(2)_R\times\widetilde{SU(2)}_R=SU(2)_b^{(1)}\times SU(2)_f^{(1)}$. Although there are two $SU(2)$ groups remaining, they cannot be both flavor symmetries as the supercharges must be singlets under flavor symmetry. Moreover, with four free real scalars with conformal dimension $\Delta=1/2$ we can build only three chiral scalars with conformal dimension $\Delta=1$. The explicit construction gives a triplet of $SU(2)_{fl}\equiv diag[SU(2)_b^{(2)}\times SU(2)_f^{(2)}]$. Thus, there is just one $SU(2)$ flavor group $SU(2)_{fl}\equiv diag[SU(2)_b^{(2)}\times SU(2)_f^{(2)}]$.

The remaining three currents $J=J_b^{(2)}-J_f^{(2)}$ in the adjoint representation of $SU(2)_{fl}$ are the superprimary of an $\cN$ superconformal multiplet. According to the table from the previous section, this multiplet contains in addition to the three currents $J$ supercurrents in the representation $(2,2;3)$ of $SU(2)_R\times\widetilde{SU(2)}_R\times SU(2)_{fl}$. These and the supercurrents $(2,2;1)$ entering the superconformal algebra (those in the stress-tensor multiplet) give all the conserved supercurrents.

Furthermore, this $\cN$ multiplet contains eighteen ``stress-tensors" in the representation $(3,1;3)+(1,3;3)$. There is another ``stress-tensor" -- the real stress-tensor (singlet under $R-$ and flavor symmetries which produces $P_\mu$ in the superconformal algebra) in the stress-tensor multiplet $\cT$. The last remaining ``stress-tensor" which is $T_b-T_f$ is superprimary. Altogether, there are $20$ ``stress-tensors" obtained by taking products of the four Majorana fermions, and by taking products of four real scalars with derivatives. $T_b$ stands for the bosonic stress-tensor which is singlet under the full global symmetry group. Similarly, $T_f$ is the singlet fermionic stress-tensor. In terms of these, the singlet stress-tensor which produces translation generators from the superconformal structure is given by $T_b+T_f$.

It remains to identify the UV $R-$symmetry group $SO(4)_r\equiv SU(2)_X\times SU(2)_Y$ in terms of the complete symmetry group $SO(4)_b\times SO(4)_f\equiv SU(2)_b^{(1)}\times SU(2)_b^{(2)}\times SU(2)_f^{(1)}\times SU(2)_f^{(2)}$ of the IR fixed point. The identification is

\begin{align}
SU(2)_X=diag[SU(2)_b^{(1)}\times SU(2)_b^{(2)}\times SU(2)_f^{(1)}],\qquad SU(2)_Y=SU(2)_f^{(2)}
\end{align}

No $U(1)$ subgroup of $SO(4)_r$ corresponding to considering the UV theory as an $\cN=2$ theory is a superconformal $U(1)_R$ because the former always contains cartans of $SU(2)_b^{(2)}\times SU(2)_f^{(2)}$, which are absent for the superconformal $U(1)_R$:

\begin{align}
r=\pm(J_b^{(1)}+J_b^{(2)}+J_f^{(1)})\pm J_f^{(2)},\qquad R=\pm J_b^{(1)}\pm J_f^{(1)}.
\end{align}

Here $r$ and $R$ are UV and IR $R-$charges, correspondingly. All $J'$s are cartans of the corresponding $SU(2)$'s.

\subsection{General analysis of (super)currents multiplets}

\begin{itemize}
\item Presence of any one of the $\cN,\cS$ multiplets implies that the theory contains a free sector when this collection of the multiplets is consistent.
\end{itemize}
Indeed, each of these multiplets contains ``stress-tensors" which are not singlets of $SO(4)_R$. The existence of additional ``stress-tensors" means that the theory is composed of several sectors. Furthermore, a non-free $\cN=4$ sector contributes only a single $\cT$-multiplet. Thus, the theory in hand must contain the free $\cN=4$ and/or the free $\cN=8$ sectors.

The consistent collection of these multiplets is then the collection given by these free theories. These collections are described in section 3.3 and Appendix B.

\begin{itemize}
\item Each non-free sector must have at least $\cN=4$ supersymmetry. The amount of (super)symmetry of a product of theories can exceed those of the products only if at least two products are free.
\end{itemize}

Indeed, an enhancement in symmetries occurs when a conserved primary current $J$ is a product of two local operators belonging to different theories $J(x)=\cO^{(1)}(x)\cO^{(2)}(x)$. This product is nonsingular exactly because the two factors belong to two different factors in the tensor product of the theories. For a product of two (super)conformal theories there are exactly three (super)conformal structures: the structure of the first factor, the structure of the second factor and the diagonal structure. We are dealing with the diagonal structure so that $\Delta(J)=\Delta^{(1)}(\cO^{(1)})+\Delta^{(1)}(\cO^{(1)})$.

Using the unitarity constraints for primary operators \cite{Minwalla}
\begin{eqnarray}
& \Delta\ge j+1,\qquad\hbox{for}\quad j>1/2,\nonumber\\
& \Delta\ge 1,\qquad\hbox{for}\quad j=1/2,\nonumber\\
& \Delta\ge\frac12,\qquad\hbox{for}\quad j=0,
\end{eqnarray}
it is trivial to show that any ``enhanced" primary conserved current $J$ whether it is a spin one current ($\Delta=2,j=1$), a supercurrent ($\Delta=5/2,j=3/2$) or a stress tensor ($\Delta=3,j=2$) is build with free fields: both $\cO^{(1)}$ and $\cO^{(2)}$ are either free or descendant of free fields. The free scalar has conformal dimension $\Delta=1/2$ and the free fermion has conformal dimension $\Delta=1$.

\begin{itemize}
\item  $\cE$ can only be present in the following cases
 \begin{itemize}
  \item without any flavor symmetries and leads to susy enhancement from $\cN=4$ to $\cN=5$.
  \item with a single $U(1)$  flavor group leading to susy enhancement from $\cN=5$ to $\cN=6$.
  \item  with $SU(2)_{fl}^2$ and implies susy enhancement $\cN=4\to\cN=8$, either free or not.
 \end{itemize}
\end{itemize}

The $\cE$ multiplet does not contain a ``stress tensor", so the SCFT is irreducible. At the same time, the four supercharges acting on additional currents with $\Delta=2,j=1$ and in fundamental representation of $SO(4)_R$ produce additional supercurrents in the singlet representation of $SO(4)_R$. In other words, the commutator of four supercharges and 4 global charges give another supercharge. This can only happen in the following cases.

(1) The supersymmetry is enhanced from $\cN=4$ to $\cN=8$. In this case there must be two more $SU(2)$ groups which are both flavor from the $\cN=4$ point of view. Under this $SU(2)_{fl}^2\equiv SO(4)_{fl}$ the superprimary of the $\cE$ supermultiplet must be in representation $(2,2)$. The $SO(4)_{fl}$ is the commutant of $SO(4)_R$ in the complete global symmetry group $SO(8)$ which is the $R-$symmetry group of $\cN=8$ SCFT.

(2) When there are no flavor symmetries (no $\cF-$supermultiplets), essentially the same argument forces enhancement of susy to $\cN=5$.

(3) Finally, when there is a single $U(1)$ flavor symmetry, susy is enhanced to $\cN=6$ with the $U(1)$ symmetry being the commutant of $SO(4)_R$ in $SO(6)_R$.

As will be shown in the next section, the last two cases of susy enhancement (to $\cN=5$ and $\cN=6$) do not happen for a ``bad" UV theory, so they are not the subject of the present paper.

\subsection{Bad $SO(4)_r$ UV symmetries}

Now that we classified all (super)currents multiplets we can proceed with the classification of bad theories.

\begin{itemize}
\item No additional (in addition to the stress-tensor multiplet $\cT$) supermultiplets from Table 1.
\end{itemize}

In this case the only conserved currents are those from the stress tensor supermultiplet. In particular, the full global symmetry group is the $R-$symmetry group $SO(4)_R$. In this case there is no room for a different UV $R-$symmetry group, so the theory is good. That is, it is irreducible and $SO(4)_r=SO(4)_R$.

\begin{itemize}
\item Additional $\cT-$ supermultiplet(s) only.
\end{itemize}

In this case the IR theory is reducible and is the product$T=T^{(1)}\otimes\ldots\otimes T^{(k)}$ of $k$ interacting $\cN=4$ theories with the superconformal structure being the diagonal one. The full global symmetry group is $SU(2)^{(1)}\times\widetilde{SU(2)^{(1)}}\times\ldots\times SU(2)^{(k)}\times\widetilde{SU(2)^{(k)}}$. The $R-$symmetry group is
\begin{eqnarray}
&SO(4)_R=SU(2)_R\times\widetilde{SU(2)}_R=\\
& diag[SU(2)^{(1)}\times\ldots\times SU(2)^{(k)}]\times diag[\widetilde{SU(2)^{(1)}}\times\ldots\times \widetilde{SU(2)^{(k)}}]\nonumber
\end{eqnarray}
 The choices of $SO(4)_r=SU(2)_X\times SU(2)_Y$ that correspond to ``bad" UV theories are
\begin{eqnarray}
& SU(2)_X=diag[SU(2)^{(i)}\times P_X]\nonumber\\
& SU(2)_Y=diag[\widetilde{SU(2)}^{(i)}\times P_Y]
\end{eqnarray}

where $1\le i\le k$ and $P_X$ and $P_Y$ are products of some of the remaining $2k-2$ $SU(2)$ groups subject to the requirement that they do not share a common $SU(2)$ factor and one is not obtained from the other by putting tilde on top of all $SU(2)$'s that do not have it and removing them from all those that do have tilde. Otherwise, they are arbitrary.

\begin{itemize}
\item Flavor supermultiplet(s)($\cF-$supermultiplet(s)) only.
\end{itemize}

Such a theory is irreducible with $\cN=4$ superconformal symmetry. In order to be the IR limit of a ``bad" theory the flavor symmetry group must be of the form $G=SU(2)^{k}\times H$ with $k\ge 1$.
Then the UV $R-$symmetry $SO(4)_r\equiv SU(2)_X\times SU(2)_Y$ is
\begin{eqnarray}
SU(2)_X=diag[SU(2)\times SU(2)_{fl}^{i_1}\times\ldots\times SU(2)_{fl}^{i_l}],\nonumber\\
SU(2)_Y=diag[\widetilde{SU(2)}\times SU(2)_{fl}^{j_1}\times\ldots\times SU(2)_{fl}^{j_m}],
\end{eqnarray}
where $0\le l\le k$, $0\le m\le k$, $i_m\ne j_p$ for any $m$ and $p$ and $1\le l+m$.

The simplest example is when the full global symmetry group of the IR theory is $SU(2)_R\times \widetilde{SU(2)}_R\times SU(2)_{fl}$ and
\begin{eqnarray}
SU(2)_X=SU(2),\qquad SU(2)_Y=diag[\widetilde{SU(2)}\times SU(2)_{fl}].
\end{eqnarray}

\begin{itemize}
\item $\cE-$supermultiplet(s).
\end{itemize}

This case was partially discussed in the previous section. All that remains is to discuss embeddings of $SO(4)_r$ into $SO(8)_R$ which correspond to bad UV theories and show that no such embedding exist for $\cN=5$ and $\cN=6$ theories.

Start with embeddings of $SO(4)_r$ in $SO(5)$ which correspond to susy enhancement to $\cN=5$. All embeddings pick up an $SO(4)$ subgroup of $SO(5)_R$ which is an $SO(4)_R$ group of the $\cN=4$ superconformal substructure of $\cN=5$ IR SCFT. Hence, the UV theory cannot be bad.

There is an embedding of $SO(4)$ in $SO(6)_R$ which does not correspond to an $\cN=4$ superconformal substructure of the IR $\cN=6$ SCFT, namely the two $SU(2)$ factors in $SO(4)$ can be identified with the obvious subgroup $SO(3)\times SO(3)$ of $SO(6)_R$. However, with respect to $SO(3)\times SO(3)\subset SO(6)_R$ the supercharges form the representation $(3,1)+(1,3)$, while for an $\cN=4$ UV structure we need supercharges in representation $(2,2)$. Thus, this case does not work, either.

As there are no $\cN=7$ SCFT's except those that are $\cN=8$ \cite{B2}, the remaining option is embeddings of $SO(4)_r$ in $SO(8)_R$. Appropriate embeddings (corresponding to ``bad" UV theories) exist thanks to the existence of $SU(2)^{n>2}$ subgroup of $SO(8)$ group. There is a natural $SO(4)\times SO(4)$ subgroup of $SO(8)$ which is equivalent to $SU(2)_a\times SU(2)_b\times SU(2)_c\times SU(2)_d$ with respect to which the adjoint ${\bf 28}$, the vector ${\bf 8}_v$ and two spinor representations ${\bf 8}_c$ and ${\bf 8}_s$ decompose as follows
\begin{eqnarray}
& {\bf 28}=(3,1,1,1)+(1,3,1,1)+(1,1,3,1)+(1,1,1,3)+(2,2,2,2),\nonumber\\
& {\bf 8}_v=(2,2,1,1)+(1,1,2,2),\nonumber\\
& {\bf 8}_c=(2,1,2,1)+(1,2,1,2),\nonumber\\
& {\bf 8}_s=(2,1,1,2)+(1,2,2,1.)
\end{eqnarray}

Different ways to embed $SO(4)_r$ into $SO(8)_R$ correspond to different ways to embed $SO(4)_r$ into $SU(2)^4$. There are three distinct patterns for forming the UV $R-$symmetry group $SO(4)_r\equiv SU(2)_X\times SU(2)_Y$

\begin{align}
& (i)\qquad SU(2)_X=SU(2)_a,\nonumber\\
& \quad SU(2)_Y=diag[SU(2)_b\times SU(2)_c];\\
& (ii)\qquad SU(2)_X=SU(2)_a,\nonumber\\
& SU(2)_Y=diag[SU(2)_b\times SU(2)_c\times SU(2)_d];\\
& (iii)\qquad SU(2)_X=diag[SU(2)_a\times SU(2)_c],\nonumber\\
& SU(2)_Y=diag[SU(2)_b\times SU(2)_d].
\end{align}

They all correspond to bad UV theories.

\begin{itemize}
\item $\cN-$supermultiplet.
\end{itemize}

The only theories that have $\cN-$supermultiplets are the free $\cN=4$ and $\cN=8$ SCFT's and their products with arbitrary $\cN=4$ SCFT's. In a single $\cN=4$ free SCFT $\cN-$supermultiplets for a triplet under the flavor symmetry group $SU(2)_{fl}$. In a single $\cN=8$ free SCFT $\cN-$supermultiplets are in the representation $(1,1;1,3)+(1,1;3,1)$ of $SO(4)_R\times SO(4)_{fl}\subset SO(8)_R$ (see Appendix B).

\begin{itemize}
\item $\Psi-$supermultiplet.
\end{itemize}

The only $\cN=4$ SCFT that posses this supermultiplet is the free $\cN=8$ superconformal theory (or any reducible $\cN=4$ SCFT, at least one of whose sectors is the free $\cN=8$ SCFT.). Although this supermultiplet does not contain additional ``stress-tensors", it has additional spin-1 and supercurrents which can only be consistently embedded into the free $\cN=8$ SCFT. See appendix B for the list of (super)current supermultiplets of this theory.

In other words, if such a multiplet exist in an $\cN=4$ SCFT, then there must also exist the flavor multiplets corresponding to the flavor symmetry $SU(2)^2$ under which the superprimary spinor must transform in representation $(2,2)$. The $SU(2)^4$ global symmetry is enhanced to $SO(8)_b\times SO(8)_f$ and all the multiplets from Appendix B are present.

\begin{itemize}
\item $\cS-$supermultiplet.
\end{itemize}

This supermultiplet exists only in the free $\cN=8$ SCFT in two copies which are in the representations $(1,1;2,2)+(1,1;2,2)$ under $SO(4)_R\times SU(2)_{fl}^2$.

\begin{itemize}
\item Products of the free $\cN=4$ and $\cN=8$ SCFT's.
\end{itemize}

Such products contain supermultiplets from Table 1 with flavor quantum numbers different from those of a single free $\cN=4$ and a single $\cN=8$ SCFT's. In these cases all the flavor quantum numbers can be easily deduced with the help of Table 1 by taking products of (derivatives) of free fields.

\section{$\cN=4$ $SU(2)$ SYM with two fundamental flavors as a bad theory}

In this section we consider a genuine interacting $\cN=4$ SCFT which is the IR limit of a bad UV theory -- the $\cN=4$ SYM with gauge group $SU(2)$ and two hypermultiplets in the fundamental representation of the gauge group.

First of all let us compute the UV $R-$charge of monopole operators
\begin{align}
R=-2|n|+\frac{N_f}{2}(|n|+|-n|)=0
\end{align}
where $N_f$ is the number of hypermultiplets, and $n$ is half integral corresponding to the GNO charge
\begin{align}
H=\begin{pmatrix}
n & 0\\
0 & -n
\end{pmatrix}
\end{align}

As was argued in section 2, this makes the would-be superconformal index divergent which indicates an inequality between UV and IR $R-$symmetries. Thus the theory is bad. Furthermore, as argued by Seiberg \cite{Seiberg}, this superconformal theory is interacting.

Let us consider the moduli space. The Higgs branch is two cones $H_L$ and $H_R$, each one is a copy of ${\mathbb C}^2/{\mathbb Z}_2$ with the common origin \cite{SW}. They are parameterized by gauge-invariant operators $V_L$ and $V_R$ which transform in the representations $({\bf3,1,3,1})$ and $({\bf1,3,3,1})$ of the UV symmetry group $SU(2)_{f_1}\times SU(2)_{f_2}\times SU(2)_R\times SU(2)_N$, correspondingly. Here $SU(2)_{f_1}\times SU(2)_{f_2}\equiv SO(4)_f$ is the flavor symmetry group, and $SU(2)_R\times SU(2)_N$ is the UV $R-$symmetry group. Furthermore, they satisfy relations $V_LV_L\supset({\bf1,1,5,1})=0$ and $V_RV_R\supset({\bf1,1,5,1})=0$, so that both cones have the correct real dimension four.

The metric on the four-dimensional Coulomb branch of this theory was argued in \cite{SW2} not to have quantum corrections. So it is the flat metric on $({\mathbb R}^3\times S^1)/{\mathbb Z_2}$ where ${\mathbb Z_2}$ acts by changing signs of all four coordinates. There are two fixed points of the ${\mathbb Z}_2$ action: $({\bf 0},0)$ and $({\bf 0},\pi)$. To get the Coulomb branch of an IR fixed points one takes the scaling limit of the metric. When the limit is taken at either of the singular points, the Coulomb branch becomes the cone ${\mathbb C}/{\mathbb Z}_2$.

This moduli space can be that of an interacting SCFT or the free $\cN=4$ theory with gauged discreet symmetry group ${\mathbb Z}_2$ which acts by changing signs of the elementary hypermultiplet. Employing string theory arguments for the gauge theory, Seiberg argued that the interacting SCFT option is realized at the origin of the moduli space \cite{Seiberg}.

The isometry group of the cone ${\mathbb C}^2/{\mathbb Z}_2$ is $SO(4)\equiv SU(2)\times SU(2)$ whose action is the standard action of $SO(4)$ on ${\mathbb R}^4$ preserved by the orbifolding. In order to see the relation between the $SU(2)_N$ which acts on the Coulomb branch and this complete symmetry group of the cone, it is more convenient to consider the Coulomb branch in terms of four real coordinates subject to orbifolding. In this picture $SU(2)_N$ acts on three out of four real coordinates. This is obvious when we go far away from the origin of the Coulomb branch of the gauge theory and take into account the fact that the three real scalars of the vector multiplet form a triplet of $SU(2)_N$ while the dualized photon is invariant. When $SO(4)$ is written as $SU(2)_{N_1}\times SU(2)_{N_2}$, it becomes obvious that $SU(2)_N=diag[SU(2)_{N_1}\times SU(2)_{N_2}]$.

So, the moduli space of the IR fixed point of the gauge theory is the three cones isomorphic to ${\mathbb C}^2/{\mathbb Z}_2$ which meet at a single point where their tips are. This point is where the interacting SCFT phase is. At this point the total global symmetry group $SU(2)_{f_1}\times SU(2)_{f_2}\times SU(2)_R\times SU(2)_{N_1}\times SU(2)_{N_2}$ is unbroken.

From the analysis of bad theories performed in the previous sections, we immediately conclude that the superconformal $R-$symmetry group is $SO(4)_R=SU(2)_R\times SU(2)_{N_1}$. Here the choice between $SU(2)_{N_1}$ and $SU(2)_{N_2}$ is a convention. This corresponds to the case 3 ($\cF-$ supermultiplets only) from the section 3.5. So the $\cN=4$ $SU(2)$ gauge theory with two fundamental hypermultiplets is an example of a bad theory whose IR limit is an interacting irreducible $\cN=4$ SCFT thanks to accidental conserved currents in the representation ${\bf 3}$ of $SU(2)_N$.

Now that we identified the superconformal structure we can find conformal dimensions of BPS operators. Let us start with scalars from the vector multiplet. The gauge invariant operators bilinear in the elementary scalars $tr(\phi^i\phi^j)$ form the representation ${\bf 5}$ of $SU(2)_N$, among them one BPS and one anti-BPS operators. The ${\bf 5}$ fits in the multiplet $({\bf3,3})$ of $SU(2)_{N_1}\times SU(2)_{N_2}$ as the highest-weight representation in the decomposition $({\bf3,3})={\bf 1+3+5}$. So, since the BPS operator from ${\bf5}$ has the IR $R-$charge one, its conformal dimension is also one which implies $\Delta({\bf 1,1,3,3})=1$. According to Table 1 these scalars are the lowest component of the multiplet of the $SU(2)_{N_2}$ flavor currents. Analogously, $V_L$ and $V_R$ whose conformal dimensions are also one are the lowest components of the multiplets of flavor currents $SU(2)_{f_1}$ and $SU(2)_{f_2}$, respectively.

Recall that the 'naive' conformal dimension of the single trace operator ${\bf 5}\subset tr(\phi^i\phi^j)$ was two, so this operators acquire a large anomalous dimension $\Delta-2=-1$ in the IR SCFT.

The same logic gives the conformal dimension of all single-trace operators $({\bf 2k+1})\subset tr(\phi^{i_1}...\phi^{i_k})$: $\Delta({\bf 2k+1})=k$.  Note, that although these operators are BPS, the UV description of the SCFT suggests wrong conformal dimensions. This was possible to rectify by finding the correct IR superconformal structure. A significant role was played by classification of bad theories developed in the previous sections.

Now let us turn attention to the operators in the representations ${\bf 1+3}$ of $SU(2)_N$. These are operators with conformal dimension $\Delta=1$ in IR which do not correspond to any gauge-invariant operators in the UV theory build from the elementary fields. So it is natural to look for them among the monopole operators.

Analysis of the spectrum of monopole operators in UV of $\cN=4$ super Yang-Mills theories in radial quantization was carried out in \cite{BKK}. They found that the spectrum consisted of monopole operators in the representations of $SU(2)_N$ with spins $l_{min},l_{min}+1,l_{min}+2,...$ with
\begin{align}
l_{min}=l_{min}(H)=-\frac12\sum_{\hbox{roots}{}\alpha}|\alpha(H)|+\frac{N_f}{2}\sum_{\hbox{weights}{}\rho}|\rho(H)|.
\end{align}
In this formula $H$ is the GNO charge.

For the present case $l_{min}=0$ for all GNO charges, so this description suggests that there are monopole operators in all integral spin representations of $SU(2)_N$ in the UV. This matches the IR spectrum. Indeed, we already have local operators in the representations ${\bf 1}$ and ${\bf 3}$, but there are all other integral spin representations, too. They are obtained by taking the maximal spin representation in the product of $k$ copies of ${\bf 3}$. These operators fit into the maximal spins symmetric products $({\bf 2k+1,2k+1})$ of $k$ of $({\bf 3,3})$ of $SU(2)_{N_1}\times SU(2)_{N_2}$ which also contain the maximal spin symmetric products of $k$ of ${\bf 5}\subset tr(\phi^i\phi^j)$. Thus, these operators in the representations ${\bf 2k+1}$ of $SU(2)_N$ have IR conformal dimensions $\Delta=k$.

Although we find match between monopole states in the radial quantization of the UV theory and the local operators of the IR fixed point, it is not obvious that this should be the case, as the states/operators do not even contain (anti)BPS operators.

\section{Conclusions and outlook}

In this paper we investigated the options for the structure of the IR fixed points of bad $\cN=4$ supersymmetric Quantum Field Theories in three dimensions, in particular, restrictions imposed by unitarity. It remains an open question whether all these possibilities are realized. In this respect the present work is a first step towards understanding the IR behavior of bad theories. The next step would be an identification of bad theories with IR structure different from that in already known examples (product of an interacting theory and free one(s)).

\section{Acknowledgements}

This research is supported by the Perimeter Institute for Theoretical Physics. Research at the Perimeter Institute is supported by the Government of Canada through Industry Canada and by Province of Ontario through the Ministry of Economic Development and Innovation.

\section{Appendix A}
The $\cN=8$ $U(1)$ gauge theory is a free theory in which the photon can be dualized to a compact scalar which decompactifies in the IR. So, in IR there is one more scalar in addition to those that were present in the UV in the gauge theory formulation. The fields fall into multiplets of $Spin(8)_R$ in the following way: scalars are the vector ${\bf 8}$, fermions and supercharges are in two different spinor representations. This is because, schematically, $Q=\phi\psi$, so if fermions are in ${\bf 8}_s$, then supercharges are in ${\bf 8}_c$.

However, in the superconformal structure the supercharges must be in the vector representation of $Spin(8)_R$. In a related manner, the scalars must be in a spinor representation in order for them to have the canonical conformal dimension $\Delta=1/2$ corresponding to free scalar fields. Thus we need to apply a triality transformation.

Consider the $SU(2)^4$ subgroup of $Spin(8)$ which is defined as $Spin(8)\supset Spin(4)\times Spin(4)=SU(2)_1\times SU(2)_2\times SU(2)_3\times SU(2)_4$. The triality group is the group which keeps one of the factors fixed and transposes the other three, so it is $S_3$. Choose $SU(2)_1$ to be fixed, and consider the permutation $S$ of the second and the third factors.

We get

\begin{eqnarray}
& {\bf 8}_s=(2,1,1,2)+(1,2,2,1)\xrightarrow{S}(2,1,1,2)+(1,2,2,1)={\bf 8}_s\nonumber\\
& {\bf 8}_c=(2,1,2,1)+(1,2,1,2)\xrightarrow{S}(2,2,1,1)+(1,1,2,2)={\bf 8}_v\nonumber\\
& {\bf 8}_v=(2,2,1,1)+(1,1,2,2)\xrightarrow{S}(2,1,2,1)+(1,2,1,2)={\bf 8}_c
\end{eqnarray}

Equivalently, this can be viewed as a reshuffling of the generators of $Spin(8)$, so that new generators of $SU(2)_2$ are the old ones of $SU(2)_3$, and the other way around. This explains why there is no way to get the superconformal $U(1)_R$ from the UV $R$-generators. A superconformal $R$-current is embedded in the current algebra of $Spin(8)_R$ in such a way, that its commutant inside $Spin(8)_R$ is $Spin(6)$. All these currents contain accidental currents of the vector representation ${\bf 7}$ of the UV $R-$symmetry group $Spin(7)_R$. This explains why an $R_{IR}$ is not a $U(1)_R$ of the UV in the a $\cN=2$ structure of the UV theory.

\section{Appendix B Superconformal multiplets of the free $\cN=8$ SCFT}

\begin{flushleft}
 \begin{tabular}{| l | l || l | l | l | l |}
    \hline
     & Name & $\cT$ & $\cF$ & $\cF$ & $\cE$\\
    Operators & & & & &\\ \hline\hline
    $j=0$ & $\Delta=1$ & $2\times(1,1;1,1)$ & $(3,1;3,1)$ & $(1,3;1,3)$ & $(2,2;2,2)$ \\
    $j=\frac12$ & $\Delta=\frac32$ & $\surd$ & $\surd$ & $\surd$ & $\surd$\\
    $j=1$ & $\Delta=2$ & $2\times[(3,1;1,1)+(1,3;1,1)]$ & $(1,1;3,1)$ & $(1,1;1,3)$ & $(2,2;2,2)$\\
    $j=\frac32$ & $\Delta=\frac52$ & $2\times(2,2;1,1)$ & $-$ & $-$ & $(1,1;2,2)$\\
    $j=2$ & $\Delta=3$ & $2\times(1,1;1,1)$ & $-$ & $-$ & $-$\\
    \hline
 \end{tabular}

 \begin{tabular}{| l | l || l | l | l |}
    \hline
     & Name & $\Psi$ & $\cN$  \\
    Operators & & & \\ \hline\hline
    $j=0$ & $\Delta=1$ & $-$ & $-$ \\
    $j=\frac12$ & $\Delta=\frac32$ & $(1,1;2,2)$ & $-$ \\
    $j=1$ & $\Delta=2$ &  $(2,2;2,2)$ & $(1,1;3,1)$ \\
    $j=\frac32$ & $\Delta=\frac52$ &  $(3,1;2.2)+(1,3;2,2)$ & $(2,2;3,1)$ \\
    $j=2$ & $\Delta=3$ &  $-$ & $(3,1;3,1)+(1,3;3,1)$ \\
    \hline
 \end{tabular}

 \begin{tabular}{| l | l || l | l | l |}
    \hline
     & Name & $\cN$ & $\cS$ & $\Xi$ \\
    Operators & & & &\\ \hline\hline
    $j=0$ & $\Delta=1$ & $-$ & $-$ & $-$\\
    $j=\frac12$ & $\Delta=\frac32$ & $-$ & $-$ & $-$ \\
    $j=1$ &  $\Delta=2$ & $(1,1;1,3)$ & $-$ & $-$\\
    $j=\frac32$ & $\Delta=\frac52$ & $(2,2;1,3)$ & $2\times(1,1;2,2)$ & $-$\\
    $j=2$ & $\Delta=3$ & $(3,1;1,3)+(1,3;1,3)$ & $2\times(2,2;2,2)$ & $2\times(1,1;1,1)$\\
    \hline
 \end{tabular}
\end{flushleft}
\bigskip

 The multiplicities of multiplets are indicated by ``multiplicity" $\times$ the corresponding $SO(4)_R\times SU(2)_{fl}^{2}\subset SO(8)_R$ quantum numbers. In this theory there exists a new supermultiplet which we denoted by $\Xi$ whose superconformal primary is a ``stress-tensor".

\end{document}